\documentclass[aps,floatfix,twocolumn,showpacs,epsfig]{revtex4}
\usepackage{amsmath}
\usepackage{epstopdf}
\usepackage{mathrsfs}
\usepackage{natbib}
\usepackage{dcolumn}
\usepackage{graphicx}
 \newcommand{\bq}{\begin{equation}}
 \newcommand{\eq}{\end{equation}}
 \newcommand{\bqn}{\begin{eqnarray}}
 \newcommand{\eqn}{\end{eqnarray}}

\newcommand{\be}{\begin{equation}}
\newcommand{\ee}{\end{equation}}
\newcommand{\bea}{\begin{eqnarray}}
\newcommand{\eea}{\end{eqnarray}}
\begin{document}
\title{\large \bf Asymptotically FRW black holes }

\author{J.T. Firouzjaee}
\affiliation{Department of Physics, Sharif University of Technology,
Tehran, Iran} \email{firouzjaee@physics.sharif.edu}

\author{Reza Mansouri}
\affiliation{Department of Physics, Sharif University of Technology,
Tehran, Iran and \\
  School of Astronomy, Institute for Research in Fundamental Sciences (IPM), Tehran, Iran}
 \email{mansouri@ipm.ir}

\date{\today}

\begin{abstract}

 Special solutions of the LTB family representing collapsing over-dense
regions extending asymptotically to an expanding closed, open, or
flat FRW model are found. These solutions may be considered as
representing dynamical mass condensations leading to black holes
immersed in a FRW universe. We study the dynamics of the collapsing
region, and its density profile. The question of the strength of the
central singularity and its nakedness, as well as the existence of
an apparent horizon and an event horizon is dealt with in detail,
shedding light to the notion of cosmological black holes.
Differences to the Schwarzschild black hole are addressed.
\end{abstract}
\pacs{95.30.Sf,98.80.-k, 98.62.Js, 98.65.-r}
\maketitle
\section{introduction}

Let us use the term cosmological black hole for any solution of
Einstein equations representing a collapsing overdensity region in a
cosmological background leading to an infinite density at its center
{\cite{sultana}. There have been different attempts to construct
solutions of Einstein equations representing such a collapsing
central mass. Gluing of a Schwarzschild manifold to an expanding FRW
manifold is one of the first attempts to construct such a
cosmological black hole, as done first by Einstein and Straus
{\cite{Einstein Straus}. Depending of the way the model is
constructed, one is led to
un-physical behavior of the trajectories {\cite{Baker}. \\

 Models not based on a cut-and-paste technology is much more
interesting giving more information on the behavior of the mass
condensation within a FRW universe model. The first attempt is due
to McVittie {\cite{McVittie} introducing a spacetime metric that
represents a point mass embedded in a Friedmann-Robertson-Walker
(FRW) universe.  There have been many other attempts to construct
cosmological black holes such as Nolan interior solution
{\cite{nolan m}, and Sultana-Dyer solution{\cite{sultana}, each of
them contrasting some of the features one expect from theory or observation. \\

The interest for cosmological black holes in the past has been
mainly from the theoretical side to understand concepts like black
hole, singularity, horizon, and thermodynamics of black holes
{\cite{haw-bh}. Indeed, the conventional definition of black holes
implies an asymptotically flat space-time and a global definition of
the event horizon. In practice, however, the universe is not
asymptotically flat. The need for local definition of black holes
and their horizons has led to concepts such as Hayward's trapping
horizon {\cite{Hayward94}, Ashtekar's isolated horizon
{\cite{ashtekar99}, Ashtekar and Krishnan's dynamical horizon
{\cite{ashtekar02},  and Booth and Fairhurst's
slowly evolving horizon {\cite{booth04}. \\
There are cases where both apparent and event horizon maybe possibly
defined. For example, for dynamical black holes one may define the
event horizon as the very last ray to reach future null infinity or
the light ray that divides those observer who cannot escape the
future singularity from those that can {\cite{kra-hel-BH}. Eardley
proposed the conjecture that in such cases trapped surfaces can be
deformed to get arbitrarily close to the event horizon
{\cite{Eardley98}. Numerical evidence was provided in
{\cite{Krishnan05} and later proved analytically for the Vaidya
metric {\cite{Ben-Dov}. \\

The precision cosmology has opened a new arena for questions like
cosmological black holes and their behavior. New observation of our
galactic center allow to resolve phenomena near the Schwarzschild
horizon of the central black hole {\cite {doeleman}. It is therefore
desirable to have black hole models embedded in cosmological
environment to see if there may be considerable differences to the
familiar Schwarzschild black hole. There have been also increasing
interest in the gravitational lensing by a cosmological mass
condensation such as a cluster of galaxies in a cosmological
background. The simplest cases are the Kottler and the
Einstein-Straus model {\cite{rindler}. The more complex situation is
lensing by a mass condensation within a dynamical
background.\\

Now, a widely used metric to describe the gravitational collapse of
a spherically symmetric dust cloud is the so-called
Tolman-Bondi-Lemaiˆtre (LTB) metric {\cite{LTB}. These models have
been extensively studied for the validity of the cosmic censorship
conjecture {\cite{cencorship},{\cite{joshi} and {\cite{joshi-cell}.
In particular, we know already {\cite{initial condition} that,
depending upon the initial conditions defined in terms of the
initial density and velocity profiles from which the collapse
develops, the central shell-focusing singularity at $r=0$ can be
either a black hole or a locally or globally naked singularity. We
may note however, that in all these papers a compact LTB region is
glued to the Schwarzschild metric or the FRW outer universe
{\cite{mansouri}. Therefore, the results have to be taken
cautiously: any principally existent event horizon is cut off by the
outer static or homogeneous space-time. The statement may still be
correct that in a dynamic spacetime the cosmic censorship hypotheses
is valid, as discussed in {\cite{wald98}. It is also possible to
glue two different LTB metrics to study the structure formation out
of an initial mass condensation or the formation of a galaxy with a
central black hole {\cite{kra-hel-sf} and {\cite{kra-hel-BH}. Here
again the structure of the metric outside the mass condensation is
defined by hand to match with a specific galaxy or cluster feature.
Faraoni et al have tried to change McVitte metric so that it
resemble a collapsing mass condensation. Their solution, however,
represents a singularity within a horizon embedded in a universe
filled with a non-perfect fluid where the change of the mass is not
because of the in-falling matter but the heat flow {\cite{faraoni2}.
This metric gives us no clue whatsoever about the dynamics of a
possible collapsed mass condensation. Harada et al, being interested
in the behavior of primordial black holes within cosmological models
with a varying gravitational constant, use a LTB solution to study
the evolution of a background scalar field when a black hole forms
from the collapse of dust in a flat Friedmann universe probing the
gravitational memory {\cite{harada}.\\

Our goal is to look for a model of a cosmological black hole, i.e. a
mass condensation leading to a singularity within a FRW universe
universe. In this paper we propose the models for closed-, open-,
and flat FRW universe studying their density profiles,
singularities, and horizon behaviors. There are many nontrivial
questions to be answered before understanding in detail the
differences of these cosmological black holes to the familiar
Schwarzschild ones, which are beyond the scope of this paper and are
to be dealt with in future publications.\\

 The question of singularities and the definition of a black
hole in such a dynamical environment has been subject of different
studies in the last 15 years. We review very shortly different
definitions of horizons in section II as a reference to the
properties of model solutions we propose. Some initial attempts to
model black holes within a FRW universe is introduced in section
III. Section IV is devoted to the LTB metric as the generic solution
representing a spherically symmetric ideal fluid. Section V is
devoted to different models of cosmological black holes, their
dynamics, density profile, apparent and event horizons,and
singularities. The question of strength and the nakedness of
singularities are dealt with in section VI. We then conclude in
section VII. Throughout the paper we assume $8\pi G = c = 1$.

\section{local definitions of black holes}

Standard definition of black holes {\cite{haw-bh} needs some global
assumptions such as regular predictability and asymptotic flatness.
In the cosmological context concepts of asymptotic flatness and
regular predictability have no application. This has already been
noticed by Demianski and Lasota {\cite{na73} stressing the fact that
in the cosmological context the standard global definition of black
holes using event horizons may not be used any more. Tipler
{\cite{tipler77} also present a definition of black hole in non
asymptotically flat space time, but these definition did not have
comprehensive property of black hole such as thermodynamic laws. In
the last decade the interest in a local definition of black holes
has led to four different concepts based primarily on
the concept of the apparent horizon. \\

Let us start by assuming a spacelike two surface $S$ with two normal null vectors
$\ell^{a}$ and $n^{a}$ on it.The corresponding expansions are then defined as
$\theta_{(\ell)}$,$\theta_{(n)}$. \\

\textbf{Definition} 1 {\cite {Hayward94}. A \emph{trapping horizon}
$H$ is a hypersurface in a 4-dimensional spacetime that is foliated
by 2-surfaces such that $\theta_{(\ell)}\mid_{H}=0$,
$\theta_{(n)}\mid_{H}\neq0$, and
$\pounds_{n}\theta_{(\ell)}\mid_{H}\neq0$. A trapping horizon is
called \emph{outer} if $\pounds_{n}\theta_{(\ell)}\mid_{H}<0$,
\emph{inner} if $\pounds_{n}\theta_{(\ell)}\mid_{H}>0$,
\emph{future} if $\theta_{(n)}\mid_{H}<0$, and \emph{past} if
$\theta_{(n)}\mid_{H}>0$. The most relevant case in the context of
black holes is the \emph{future outer trapping horizon}(FOTH).

\textbf{Definition} 2 {\cite{ashtekar99}. A \emph{weakly isolated horizon} is a
three-surface H such that :

 1. H is null;

 2. The expansion $\theta_{(\ell)}\mid_{H}=0$ where $\ell^{a}$, being null
and normal to the foliations $S$ of $H$;

 3.$-T^{b}_{a}\ell^{a}$  is future directed and causal;

 4. $\pounds_{\ell}\omega_{a}= 0$, where
$\omega_{a}=-n_{b}\nabla_{\underleftarrow{a}} \ell^{b}$, and the
arrow indicates a pull-back to H.

Weakly isolated horizon is a useful term to be used for
characterization of black holes not interacting with their
surroundings, and corresponds to isolated equilibrium states in
thermodynamics. These definition do not apply to cosmological mass
condensations because of their dynamical behavior.

\textbf{Definition}3 {\cite{ashtekar02}. A \emph{marginally trapped
tube} T (MTT) is a hypersurface in a 4-dimensional spacetime that is
foliated by two-surfaces $S$, called \emph{marginally trapped
surfaces}, such that $\theta_{(n)}|_{T} < 0$ and
$\theta_{(\ell)}|_{T} = 0$. MTTs have no restriction on their
signature, which is allowed to vary over the hypersurface. This is a
generalization of the familiar concept of the apparent horizon
{\cite{ashtekar02}. If a MTT is everywhere spacelike it is referred
to as a \emph{dynamical horizon}. If it is everywhere timelike it is
called a timelike membrane (TLM). In case it is everywhere null and
non-expanding then we have an isolated horizon. The apparent
horizons evolving in the our proposed models will not be everywhere
spacelike and will have a complex behavior. \\

Irrespective of different concepts related to the apparent horizon
we may still compromise on a definition of event horizon differing
principally from the apparent horizon. We follow the definition of
{\cite{kra-hel-BH} as the very last ray to reach future null
infinity or the light ray that divides those observer who cannot
escape the future singularity from those that can. We will see in
the next sections that cosmological black holes may have distinct
apparent and even horizons, in contrast to the Schwarzschild black
hole.

\section{Existing metrics representing over-densities within a cosmological background and their deficiencies}

\subsection{McVittie's solutions}

In 1933, McVittie {\cite{McVittie} found an exact solution of Einstein's equations
for a perfect fluid mimicking a black hole embedded in a cosmological background. McVittie's solutions can
be written in the form
\begin{equation}
ds^{2}=-(\frac{1-\frac{M}{2N}}{1+\frac{M}{2N}})^{2}dt^{2}+e^{\beta(t)}(1+\frac{M}{2N})^{4}(dr^{2}+h^{2}d\Omega^{2}),
\end{equation}
where $M=me^{\beta(t)/2}$ and $m$ is a constant. Functions $h(r)$ and $N(r)$ depend on a
constant $k$, and are given, respectively, by
\begin{multline}
\mbox{$h(r)=$}~
\begin{cases}
& \sinh(r)~~~k=1 ~~~~~~~~~~~~~~~~~~~~~~~~\\
& r~~~~~~~~~~k=0\\
& \sin(r)~~~~k=-1
\end{cases}\\
\mbox{$N(r)=$}~
\begin{cases}
& 2\sinh(r/2)~~~k=1 ~~~~~~~~~~~~~~~~~~~~~~~~\\
& r~~~~~~~~~~k=0\\
& 2\sin(r/2)~~~~k=-1
\end{cases}
\end{multline}
%\bqn  h(r) & =& \cases{\sinh r,
%& $k = -1$,\cr r, & $k = 0$,\cr
%\sin r, & $k = +1$,\cr}\nb\\
%N(r) & =& \cases{2 \sinh \frac{r}{2}, & $k = -1$,\cr r, & $k =
%0$,\cr 2\sin\frac{r}{2}, & $k = +1$.\cr} \eqn

This metric represents a point mass embedded into an isotropic
universe. It possesses a curvature singularity at proper radius $R =
2m$, in contrast to the Schwarzschild metric where there is a
coordinate singularity. It has been shown that this singularity is
space-like and weak{\cite{nolan}. The interpretation of the metric
in the region $R < 2m$ is also not clear {\cite{nolan}. Therefore,
the McVittie's metric is not a suitable solution of Einstein
equations to represent the collapse of a spherical mass distribution
with over-density within a cosmological setting.

\subsection{Sultana-Dyer solution}

Recently Sultana and Dyer {\cite{sultana} found an exact solution
representing a primordial cosmological black hole. It describes an
expanding event horizon in the asymptotic background of the Einstein-de
Sitter universe. The black hole is primordial in the
sense that it forms ab initio with the big bang singularity and
therefore does not represent the gravitational collapse of a matter distribution.\\
 This metric is given by
\begin{equation}
 ds^{2}=t^{4}[(1-\frac{2m}{r})dt^{2}-\frac{4m}{r}dtdr-(1-\frac{2m}{r})dr^{2}-r^{2}d\Omega^{2}].
\end{equation}

Though the metric has the same causal characteristics as the Schwarzschild
spacetime, there are significant differences for timelike geodesics. In
particular an increase in the perihelion precession and the
non-existence of circular timelike orbits should be mentioned. The matter content
is described by a non-comoving two-fluid source, one of which is a dust and
the other is a null fluid. At late times the dust becomes superluminal
near horizon violating the energy condition.

\section{Introducing realistic Models of Cosmological Mass Condensation}

There maybe different ways of constructing solutions of Einstein
equations representing a collapsing mass concentration in a FRW
background, as the preceding sections show. We choose the direct way
of a cosmological spherical symmetric isotropic solution, and look
for an overdensity mass distribution within the model universe
undergoing a collapse to see if and how a singularity representing a
black hole emerges. To begin with, we choose a so-called flat LTB
metric. This is the simplest spherically symmetric solution of
Einstein equations representing an inhomogeneous dust distribution
{\cite{LTB}.

\subsection{LTB metric }

 The LTB metric may be written in synchronous coordinates as
\begin{equation}
 ds^{2}=dt^{2}-\frac{R'^{2}}{1+f(r)}dr^{2}-R(t,r)^{2}d\Omega^{2}.
\end{equation}
It represents a pressure-less perfect fluid satisfying
\begin{eqnarray}\label{ltbe00}
\rho(r,t)=\frac{2M'(r)}{ R^{2}
R'},\hspace{.8cm}\dot{R}^{2}=f+\frac{2M}{R}.
\end{eqnarray}
Here dot and prime denote partial derivatives with respect to the
parameters $t$ and $r$ respectively. The angular distance $R$,
depending on the value of $f$, is given by
\begin{eqnarray}\label{ltbe1}
R=-\frac{M}{f}(1-\cos \eta(r,t)),\nonumber\\
\hspace{.8cm}\eta-\sin \eta=\frac{(-f)^{3/2}}{M}(t-t_{n}(r)),
\end{eqnarray}
\begin{eqnarray}\label{dotltbe1}
\dot{R}=(-f)^{1/2}\frac{sin(\eta)}{1-cos\eta},
\end{eqnarray}
for $f < 0$,
 and
\begin{equation}\label{ltbe2}
R=(\frac{9}{4}M)^{\frac{1}{3}}(t-t_{n})^{\frac{2}{3}},
\end{equation}
 for $f = 0$, and
\begin{eqnarray} \label{ltbe3}
R=\frac{M}{f}(\cosh \eta(r,t)-1),\nonumber\\
\hspace{.8cm}\sinh \eta-\eta=\frac{f^{3/2}}{M}(t-t_{n}(r)),
\end{eqnarray}
for $f > 0$.\\
The metric is covariant under the rescaling
$r\rightarrow\tilde{r}(r)$. Therefore, one can fix one of the three
free parameter of the metric, i.e. $t_{n}(r)$, $f(r)$, and $M(r)$.
The function $M(r)$ corresponds to the Misner-Sharp mass in general
relativity, as shown in the general case of spherically symmetric
solutions of Einstein equations{\cite{mis-sharp}.  \\
There are two generic singularities of this metric: the shell
focusing singularity at $R(t,r)=0$, and the shell crossing one at
$R'(t,r)=0$. To get rid of the complexity of the shell focusing
singularity, corresponding to a non-simultaneous big bang
singularity, we will assume $t_{n}(r)=0$. This will enable us to
concentrate on the behavior of the collapse of an overdensity region
in an expanding universe without interfering with the complexity of
the inherent bang singularity of the metric. \\
 Now, an expanding universe means generally $\dot{R}>0$.
However, in a region around the center it may happen that
$\dot{R}<0$, corresponding to the collapsing region. It is then easy
to show that in this collapsing region
$\theta_{(\ell)}\propto(1-\frac{\sqrt{\frac{2M}{R}+f}}{\sqrt{1+f}})$,
$\theta_{(n)}\propto(-1-\frac{\sqrt{\frac{2M}{R}+f}}{\sqrt{1+f}})<0$.
Therefore, $R = 2M$, is obviously a \emph{marginally trapped tube},
as defined in section 2, representing an apparent horizon according
to the familiar definitions{\cite{haw-bh,ashtekar02}. It will turn
out that this apparent horizon is not always spacelike and can have
a complicated behavior for different $r$, as was first seen in
{\cite{booth-mtt}.

\subsection{Behavior of the curvature function $f(r)$}

Now, we are interested in an expanding universe, meaning generally
$\dot{R}>0$. However, in a region around the center we expect to
have a late time behavior $\dot{R} < 0$ corresponding to the
collapse phase of the overdensity region. From equations
(\ref{ltbe1}), (\ref{ltbe2}), and (\ref{ltbe3}) we infer that to
have a collapsing region one has to ask for $f(r)<0$ in that region.
In contrast, the universe outside the collapsing region being
expanding leave us to choose $f(r)>0$, $f(r)=0$, or $f(r)<0$
depending on the model. We may have an asymptotically flat FRW
universe, however, with $f(r)>0$ or $f(r)<0$ tending to zero for large $r$. \\

 Now, we have still to make a choice for $f(r)$ at the
center $r = 0$. Expecting the mass $M$ to be zero at $r=0$ to avoid
a central singularity, we see from (\ref{ltbe1}) and (\ref{ltbe3})
that $\frac{f^{3/2}}{M}|_{r=0}=const$, or $f(r=0)=0$ {\cite{origin
con}. This give us different possibilities of the function $f(r)$ to
behave as shown in Fig.\ref{f(r)}.\\

\begin{figure}[h]
\includegraphics[scale = 0.34]{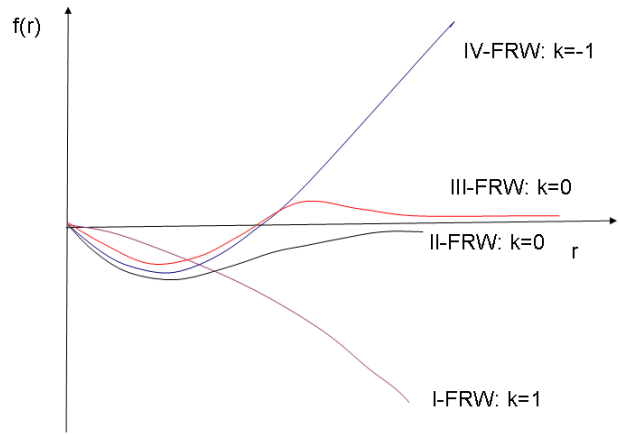}
\hspace*{10mm} \caption{ \label{f(r)}
  Different behaviors of the curvature function $f(r)$.}
\end{figure}

\section{Construction of Models}

We have now the necessary prerequisites to construct our models of
mass condensation immersed in FRW models leading to singularities
and representing cosmological black holes. Our cosmological black
hole solutions evolve from mass condensations within closed, open,
or flat FRW universes, leading to singularities having different
horizons, and providing us examples of collapsed regions behaving
differently to known Schwarzschild ones.

\subsection{Example I: $f < 0$: asymptotically closed LTB metric}

As mentioned before, we are free to choose one of the three
parameters of the LTB metric. Assuming a negative $f(r)$, we may
choose $r$ such that $f(r) = -M(r)/r$ {\cite{cencorship}. Now, let
us choose the mass function $M$ such that
$$M(r)=2^{3}a^{2}r^{3}\frac{\alpha + r^{3}}{1+r^{3}},$$
where $a$ and $\alpha$ are constants to be defined properly. We then obtain from (\ref{ltbe1})
\begin{eqnarray}
R=r(1-cos \eta(r,t))\nonumber\\
\hspace{.8cm}\eta-sin(\eta)=\sqrt{\frac{7.2+r^{3}}{1+r^{3}}}2^{3/2}at.
\end{eqnarray}

We are free to fix $a$ and $\alpha$ such that for the present time,
$t_0$, the region around the center of the overdensity, $r = 0$, is
collapsing while far from the center the universe expands. Note that
in contrast to the familiar FRW universe, where the scale factor as
a function of time, $t$, is an explicit function having a
straightforward behavior. In the LTB case, $R(t)$ playing the role
of the scale factor is an implicit function of time and comoving
coordinate $r$ given by (\ref{ltbe1}). We now fix $a$ and $\alpha$
such that $r=0$ corresponds to $\eta = \frac{3\pi}{2}$, and $r \gg
1$ corresponds to $\eta = \frac{5\pi}{6}$  (Fig. \ref{cauchy}). We
then find $a \simeq \frac{0.75}{t_{0}}$ with $t_{0}$ being the
present time, and $\alpha \simeq 7$. \\
Now, the expansion phase of the model is given by $\dot R$
(\ref{dotltbe1}). We then see from (\ref{dotltbe1}) that the region
around $r = 0$, corresponding to $\eta \sim \frac{3\pi}{2}$, is
always collapsing for any time $t$, while the regions far from the
center, $r >> 0$, at the present time, corresponding to $\eta \sim
\frac{5\pi}{6}$, are expanding. Note that this bound LTB model,
similar to the closed FRW one, has a maximum comoving
radius corresponding to $f(r) = -1$. \\
\begin{figure}[h]
\includegraphics[scale = 0.32]{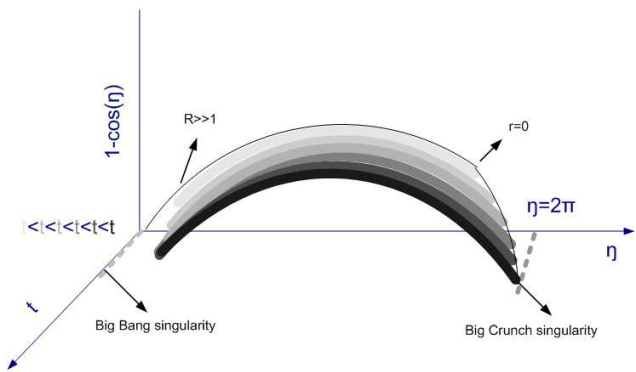}
\hspace*{10mm} \caption{ \label{cauchy}
  Evolution of the Cauchy surfaces.}
\end{figure}

The density evolution and the causal structure of the model is shown
in Fig. \ref{den1}. We see clearly how the central overdensity
region collapses to a singularity at $r = 0$, while the universe is
expanding. Note also how the slope of outgoing null geodesics tend
to infinity in the vicinity of the singularity, i.e. $ R'\rightarrow
+ \infty $ at $R = 0$.

\begin{figure}[h]
\includegraphics[scale = 0.39]{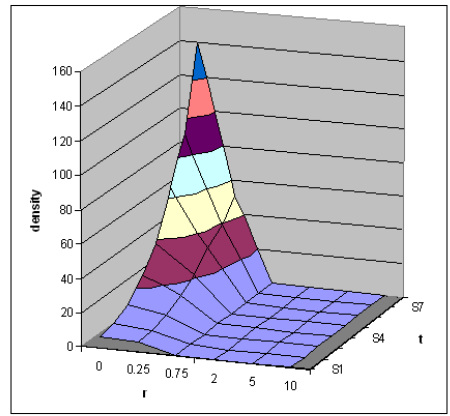}
\hspace*{10mm}
\includegraphics[scale = 0.34]{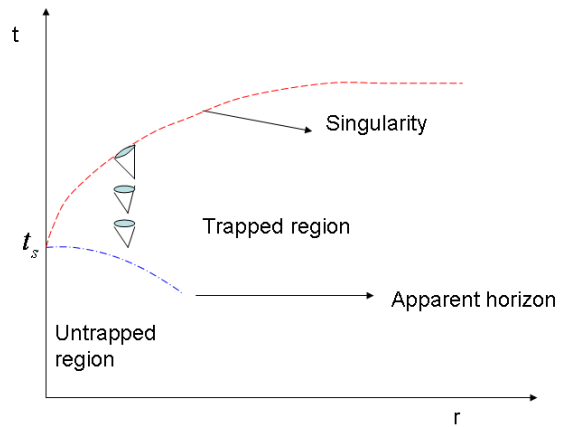}
\caption{ \label{den1} The case of the asymptotically closed
universe: in the central region the density increases with time
indefinitely while far from the center the density is decreasing
with time. The apparent horizon and the trapped region is shown in
the lower diagram.}
\end{figure}
%\begin{equation}
% a
%\end{equation}

\subsection{Example II: $f < 0$, $\lim_{r\rightarrow\infty}f(r)\rightarrow 0$; asymptotically flat LTB metric 1}

Our favorite choice is a solution representing a collapsing
overdensity region at the center and a flat FRW far from the
overdensity region. Of course the overdensity region may take part
in the expansion of the universe at early times but gradually
reversing the expansion and start collapsing. To achieve this, we
require $f(r)<0$ and $f(r)\rightarrow 0$ when $r\rightarrow\infty$.
This choice give us trivially $M(0)=0$. \\
Let us now make the ansatz $f(r)=-re^{-r}$ leading to
$$M(r)=\frac{1}{a}r^{3/2}(1+r^{3/2}),$$ where $a$ is a constant having
the dimension $[a]=[L]^{-2}$. We fix $a$ by $at_{0}=3 \pi /2$.
Similar to our previous model I, this value of $a$ corresponds to
the collapsing mass condensation around $r =0$ starting in the
expanding phase of the bound LTB model. \\

 Equation (\ref{ltbe1}),
(\ref{ltbe2}) then leads to

\begin{eqnarray}\label{met2-1}
R=\frac{\sqrt{r}(1+r^{3/2})}{a e^{-r}}(1-\cos \eta(r,t)),\nonumber\\
\hspace{.8cm}\eta-sin(\eta)=\frac{e^{-\frac{3}{2}r}}{(1+r^{3/2})}at.
\end{eqnarray}

We have plotted the density evolution and casual structure of this
model in Fig.\ref{den2}.

\begin{figure}[h]
\includegraphics[scale = 0.31]{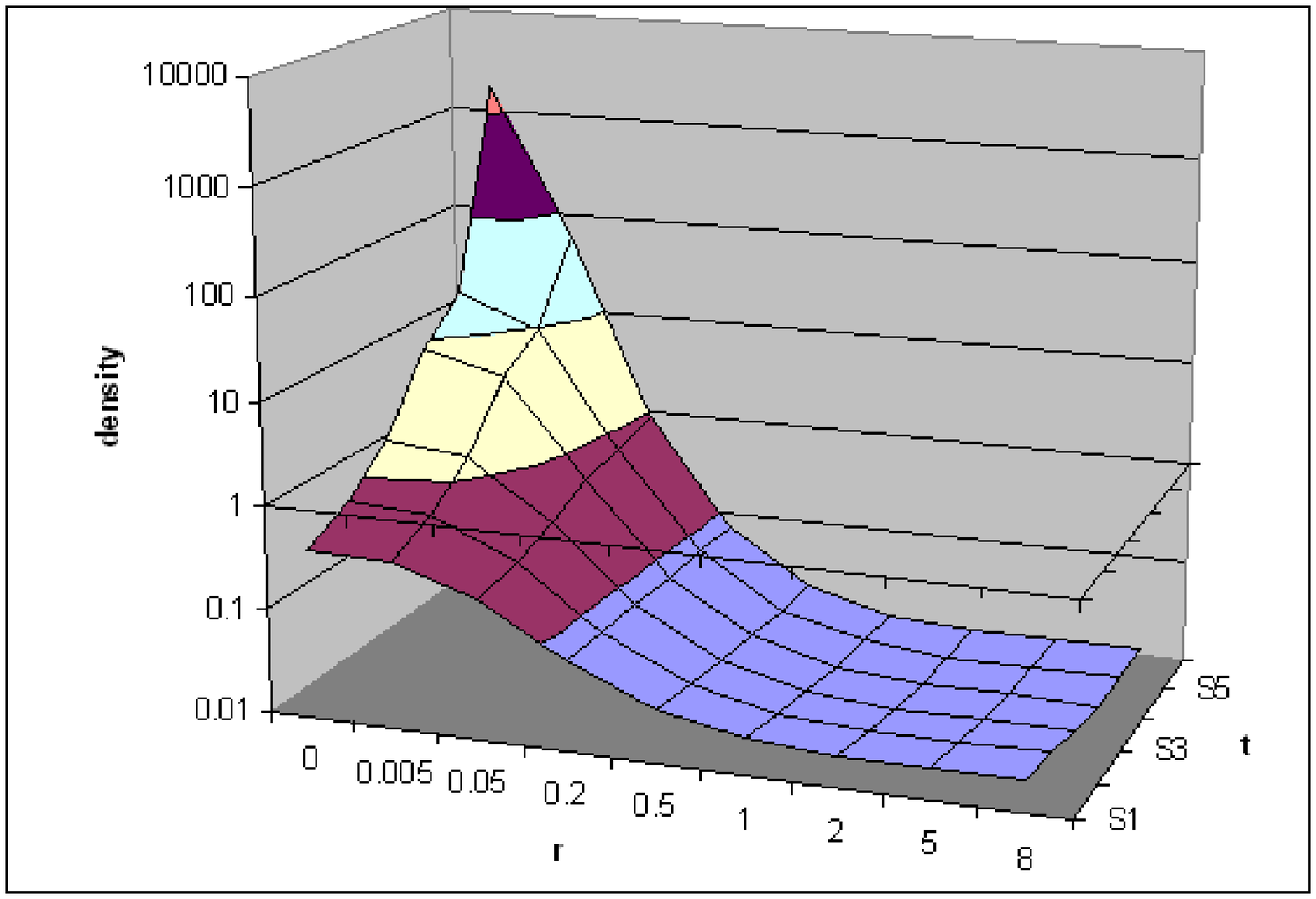}
\hspace*{10mm}
\includegraphics[scale = 0.34]{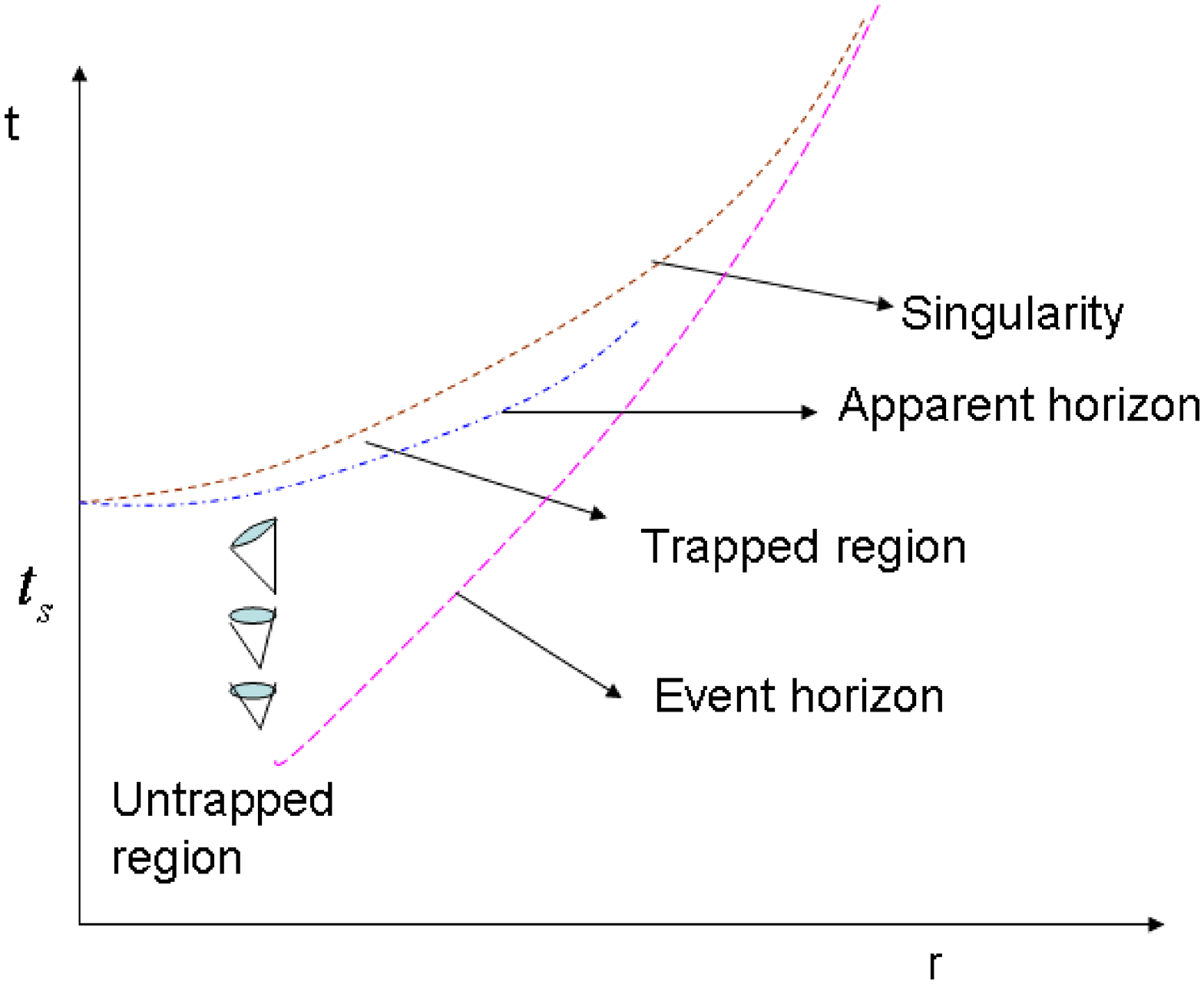}
\caption{ \label{den2}
  The density profile for the cosmic black hole within a closed but asymptotically
  flat universe. The causal structure is shown below. Note the behavior of the event
  horizon for arbitrary large but finite $t$.}
\end{figure}

As a result of $ R'\rightarrow + \infty $ near the singularity, the
slope of the outgoing null geodesics becomes infinite at the central
singularity. Again we see clearly how the collapse of the central
region and the evolution of the apparent horizon separates the
overdense region from the expanding universe.\\
The negativity of the curvature function $f(r)$ means that, although
the universe is asymptotically flat, waiting enough, every slice $r
= constant$ will collapse to the central region. We may , however,
define an event horizon according to the definition of section 2 for
any large but finite time, as shown in Fig.\ref{den2}.

\subsection{Example III: $f > 0$, $f(r)\rightarrow 0$ when $r\rightarrow\infty$; asymptotically flat LTB metric 2}

What would happen if we choose the curvature function $f(r)$ such
that it tends to zero for large $r$ while it is positive? We still
have a model which tends to a flat FRW at large distances from the
center, but having a density less than the critical one. \\
Let us make the ansatz $f(r)=-r(e^{-r}-\frac{1}{r^{n}+c})$ with $n =
2$ and $c=20000$, leading to
$$M(r)=\frac{1}{a}r^{3/2}(1+r^{3/2}),$$ where $a$ is a constant having
the dimension $[a]=[L]^{-2}$. We fix $a$ by requiring $at_{0}=3 \pi
/2$. Equation (\ref{ltbe1}), (\ref{ltbe2}) then leads to

\begin{eqnarray}\label{met2-2}
R=\frac{\sqrt{r}(1+r^{3/2})}{a (e^{-r}-\frac{1}{r^{2}+20000})}(1-\cos\eta(r,t)),\nonumber\\
\hspace{.8cm}\eta-sin\eta=\frac{(e^{-r}-\frac{1}{r^{2}+20000})^{1.5}}{(1+r^{3/2})}at.
\end{eqnarray}
and for $f>0$ region,
\begin{eqnarray}\label{met2-2}
R=\frac{\sqrt{r}(1+r^{3/2})}{a (\frac{1}{r^{2}+20000}-e^{-r})}(\cosh \eta(r,t)-1),\nonumber\\
\hspace{.8cm}\eta-sinh\eta=\frac{(\frac{1}{r^{2}+20000}-e^{-r})^{1.5}}{(1+r^{3/2})}at.
\end{eqnarray}

The solution is continuous at $r = 1$, as can be checked by
evaluating $\dot{R}$, $R'$, $\dot{R}'$, $\ddot{R}$, and $\ddot{R}'$
at $r = 1$(see the appendix).

We have plotted the density evolution and casual structure of this
model in Fig.\ref{den22}.

\begin{figure}[h]
\includegraphics[scale = 0.31]{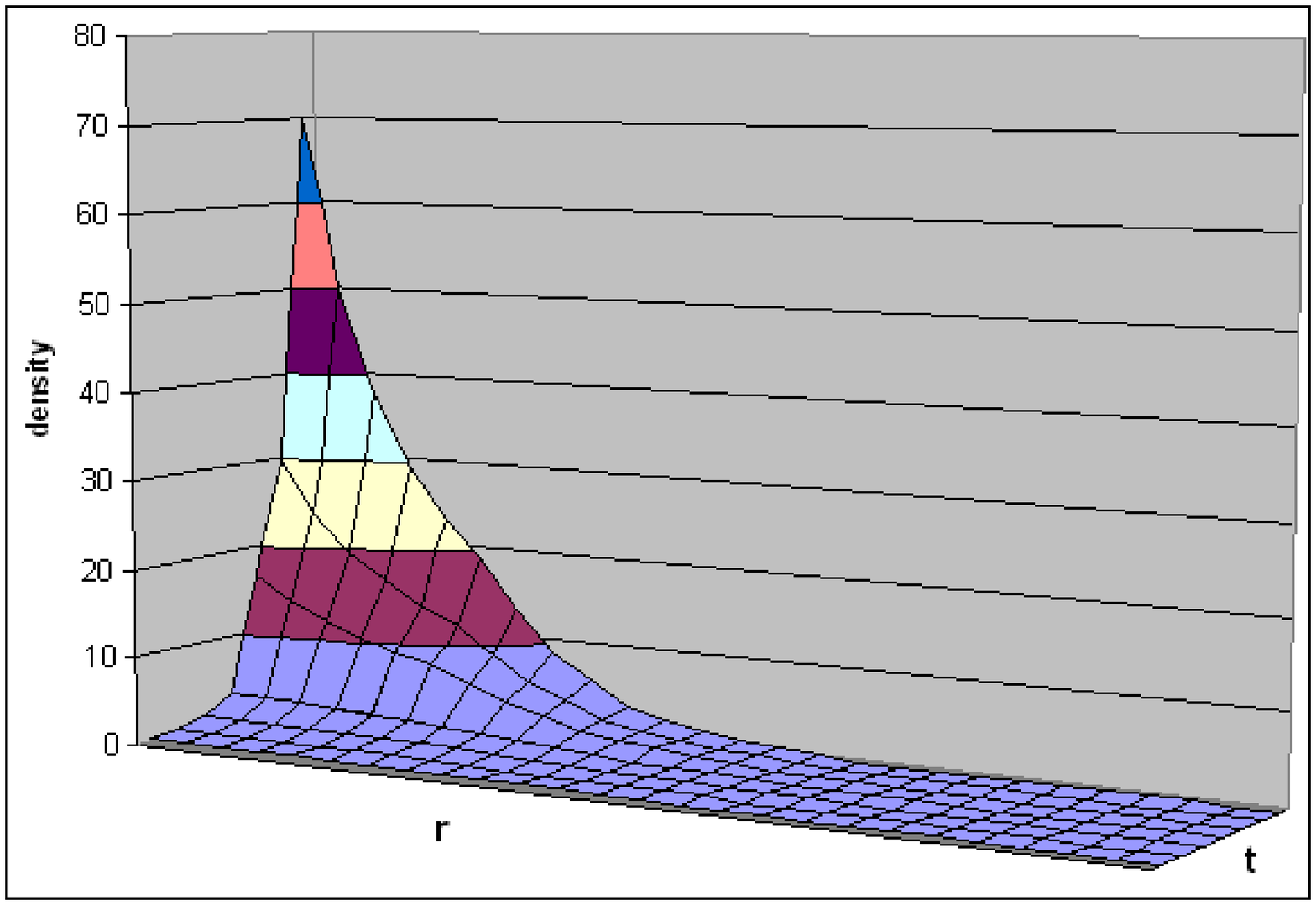}
\hspace*{10mm}
\includegraphics[scale = 0.34]{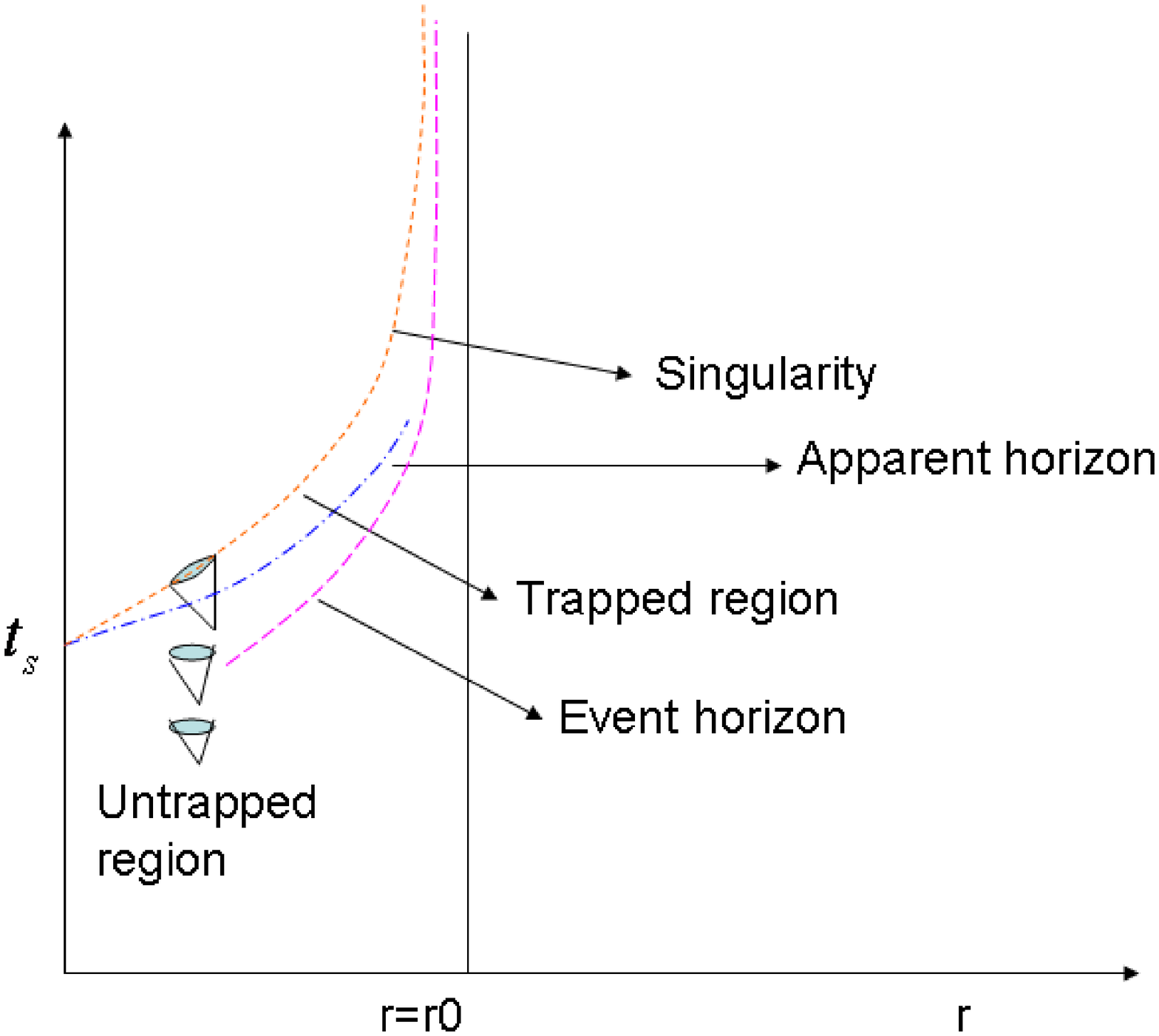}
\includegraphics[scale = 0.31]{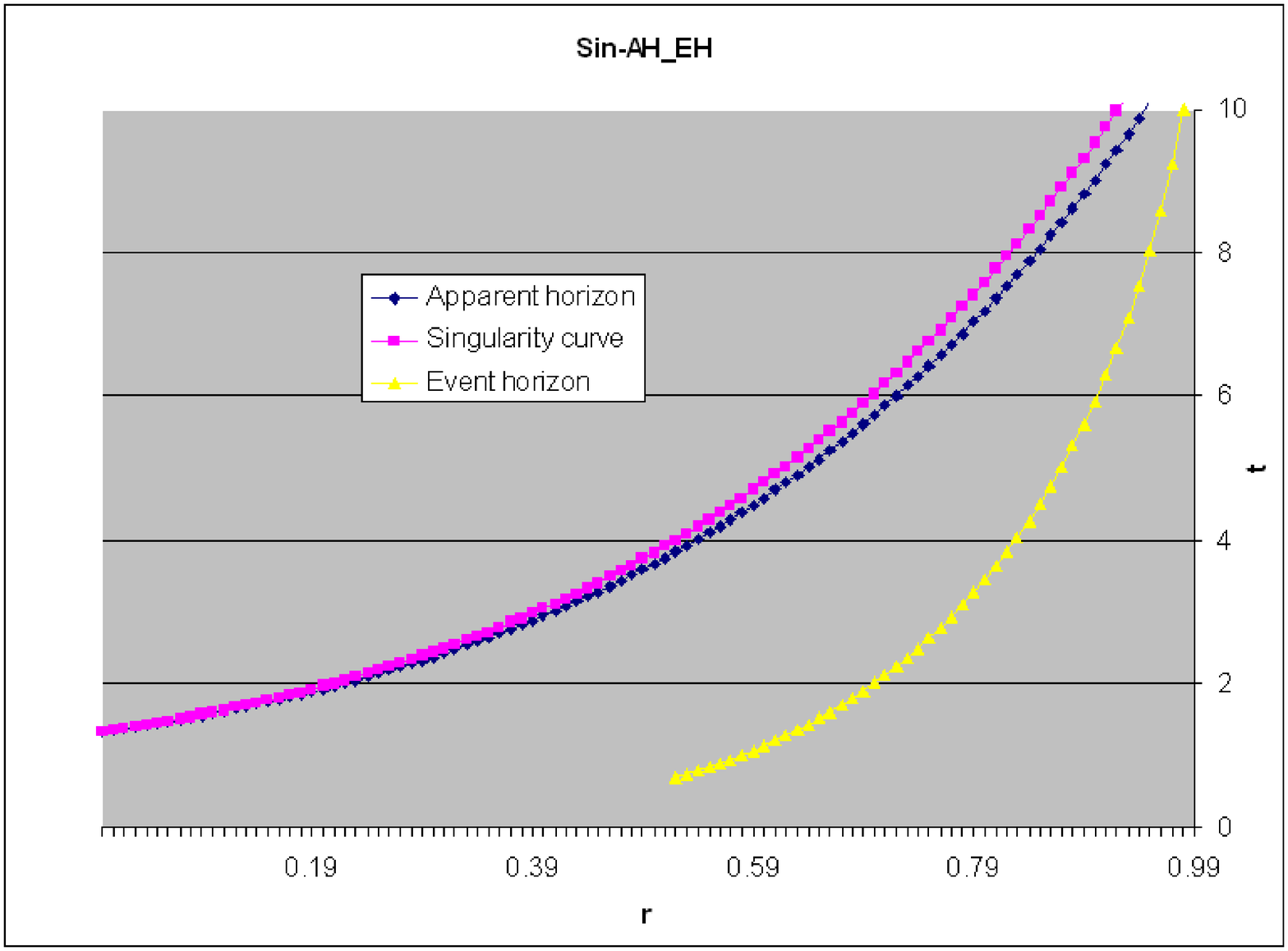}
\caption{ \label{den22}
  Evolution of the cosmic black hole within an open but asymptotically flat
  universe is similar to the closed case. The causal structure, however, is
  significantly different, as seen from the lower diagram. Result of the numerical
  calculation of the locations of the event horizon, apparent horizon and
  the singularity is also shown.}
\end{figure}

The term $\frac{1}{r^{n}+c}$ is responsible for $f(r)$ being
positive and tending to zero for large $r$ given $n\geq 2$ and
$c>>1$. Let us check if this may cause shell crossing in the region
where $f'(r)<0$ while $f>0$. Using (\ref {ltbe3}) we obtain
\begin{equation}
\frac{R'}{R}=\frac{M'}{M}(1-\phi_{4})+\frac{f'}{f}(\frac{3}{2}\phi_{4}-1),
\end{equation}
where $\frac{2}{3}\leq \phi_{4}=\frac{sinh\eta
(sinh\eta-\eta)}{(cosh\eta-1)^{2}}\leq1$. The condition for no shell
crossing singularity is then $\frac{M\mid
f'\mid}{fM'}<\frac{1-\phi_{4}}{\frac{3}{2}\phi_{4}-1}$. For
$\phi_{4}\sim 1$, corresponding to $\eta>>1$ or $t>>1$ the
inequality breaks down leading to a shell crossing singularity. The
shell crossing, however, can be shifted to arbitrary large $t$ by
choosing $f'<<1$ corresponding to $n>>1$ and $c>>1$
\cite{hell-shell}. Therefore, for the model we are proposing the
shell crossing will happen out of the range of applicability of
it.\\
As a result of $ R'\rightarrow + \infty $ near the singularity, the
slope of the outgoing null geodesics become infinite at the central
singularity. Again we see clearly how the collapse of the central
region and the evolution of the apparent horizon separates the
overdense region from the expanding universe. There is an event
horizon defined by the very last ray to reach future null infinity
and separates those observer who can not scape the future
singularity from those that can. A fixed $r = r_0$ value, being the
non-trivial root of $f(r)= 0$, divides the absolute collapsing
region from the absolute expanding region. We may be living in a
region inside the event horizon but outside the apparent one without
noticing it soon!

This solution represents a collapsing mass within an asymptotically
flat FRW universe. The collapsed region is dynamical in the sense
that its mass is not constant. In fact the rate of change of the
Misner-Sharp energy  is given by
$\frac{dM(r)}{dt}|_{R=const}=\frac{dM(r)}{dr}\frac{dr}{dt}|_{R=const}>0$
because $\frac{dM(r)}{dr}>0$, $R'dr+\dot{R}dt=0$, $R'>0$, and
$\dot{R}<0$ for collapsing region, so $\frac{dr}{dt}|_{R=const}>0$.
Therefore, it is clear that concepts such as isolated horizon and
slowly evolving horizon do not apply to this case.\\

\subsection{Example IV: $f > 0$: asymptotically open FRW metric}

Now we look for a solution which goes to an open FRW metric at
distances far from the center. At the same time one should take care
of the conditions $M(0)=0$ and $\frac{f(0)^{3/2}}{M(0)}\neq\infty$.
Let us choose $$f(r)=-r(1-r),$$ and
$$M(r)=\frac{1}{a}r^{3/2}(1+r^{3/2}),$$ where $a$ is a constant, which may be fixed by assuming $r=0$
at the present time $t_0$ corresponding to $\eta = \frac{3\pi}{2}$.
This leads to $at_{0}=3 \pi /2 +1$ . We then obtain from
(\ref{ltbe3})
\begin{eqnarray}\label{met3}
R=\frac{\sqrt{r}(1+r^{3/2})}{a(1-r)}(1-\cos\eta(r,t)),\nonumber\\
\hspace{.8cm}\eta-sin(\eta)=\frac{(1-r)^{3/2}}{(1+r^{3/2})}at,
\end{eqnarray}
for $r < 1$, and
\begin{eqnarray}
R=\frac{\sqrt{r}(1+r^{3/2})}{a(r-1)}(cosh\eta(r,t)-1)\nonumber\\
\hspace{.8cm}sinh\eta-\eta=\frac{(r-1)^{3/2}}{(1+r^{3/2})}at,
\end{eqnarray}
for $r > 1$. The solution is again continuous at $r = 1$, as can be
checked by evaluating $\dot{R}$, $R'$, $\dot{R}'$, $\ddot{R}$, and
$\ddot{R}'$ at $r = 1$(see the appendix).

 The resulting density profile and the causal structure
is plotted in Fig.\ref{den3}. Obviously a singularity at the origin
forms gradually while the universe is expanding. The causal
structure is also similar to the open but asymptotically flat case.

\begin{figure}[h]
\includegraphics[scale = 0.31]{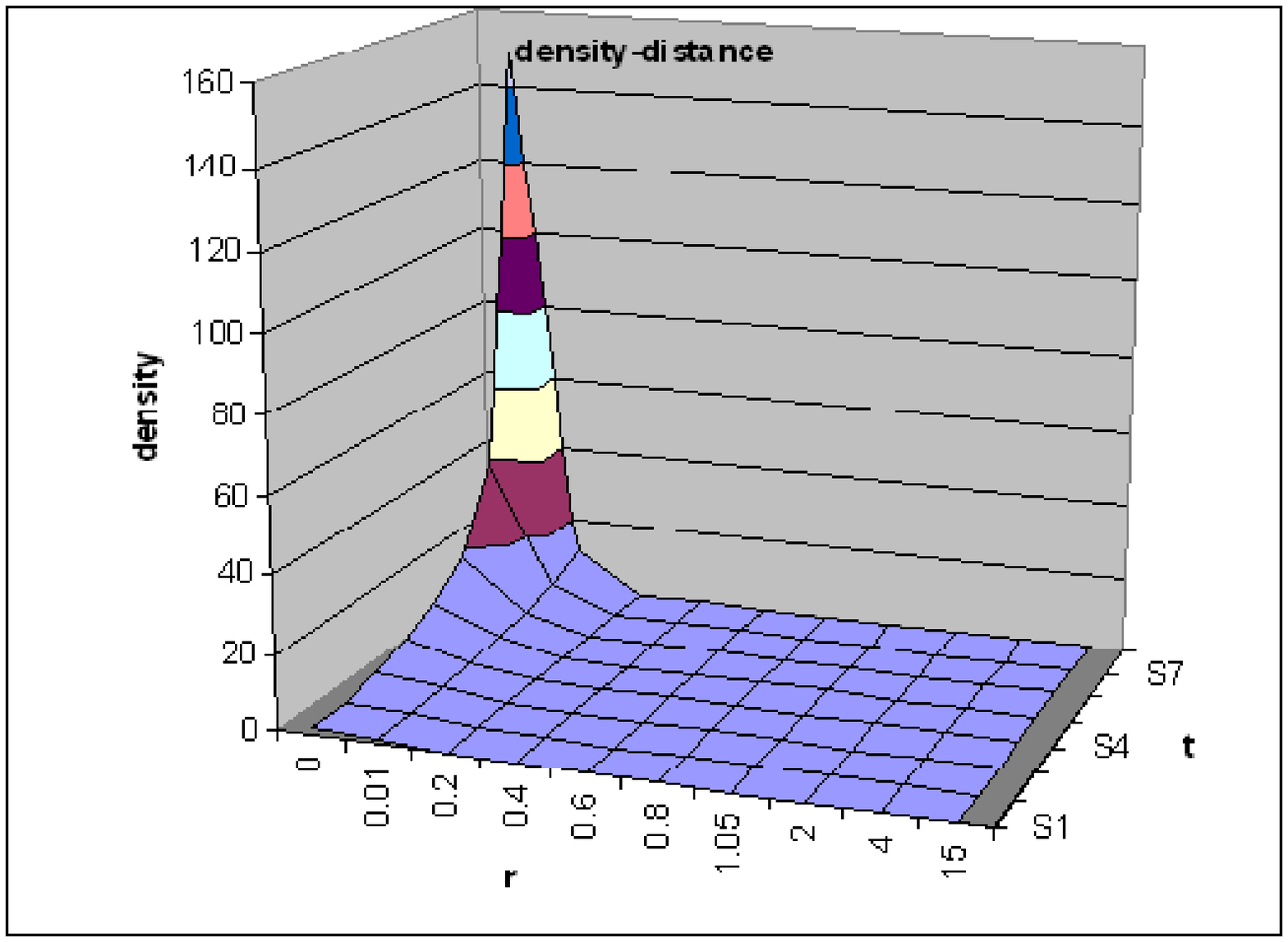}
\hspace*{10mm}
\includegraphics[scale = 0.34]{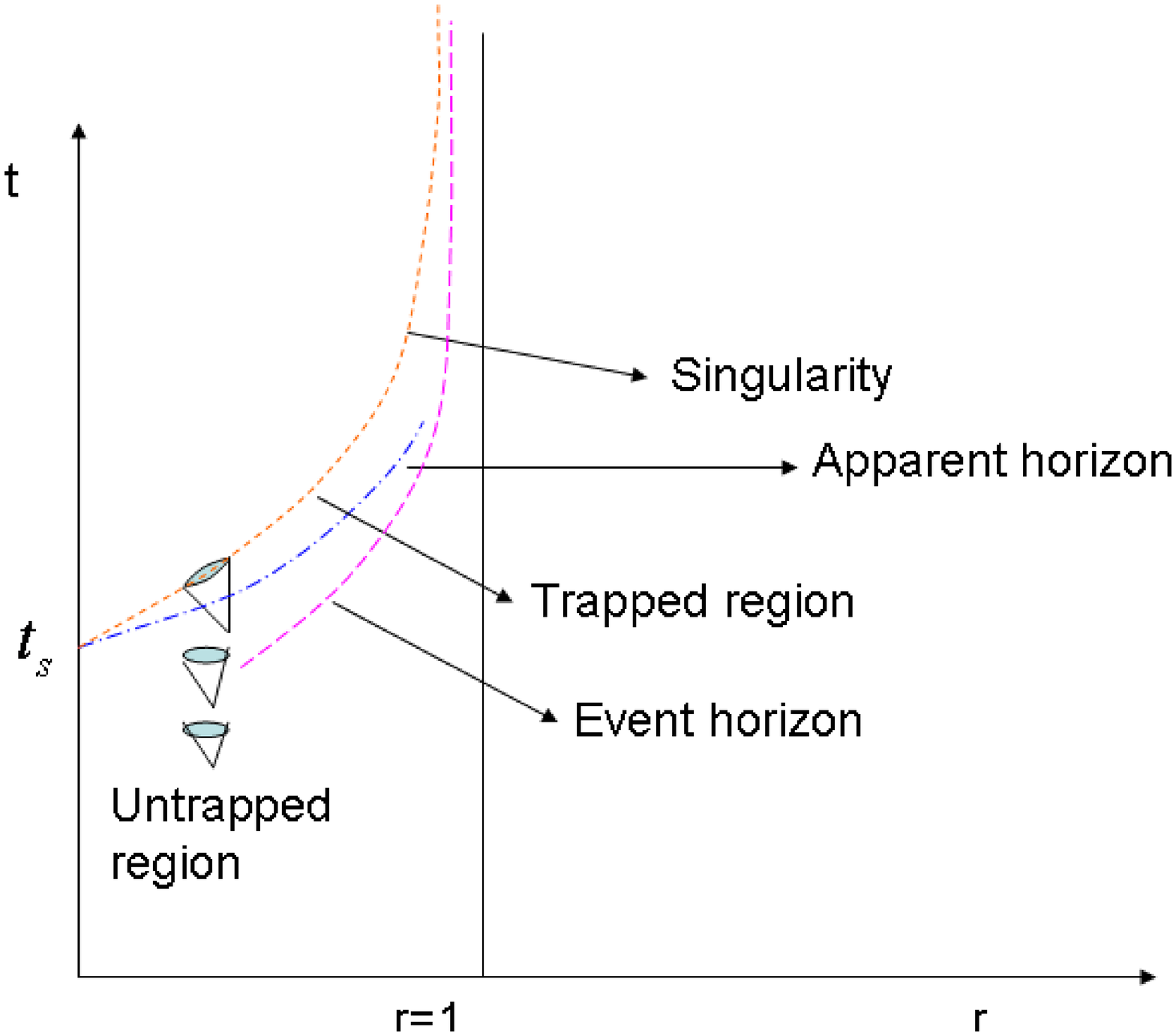}
\includegraphics[scale = 0.31]{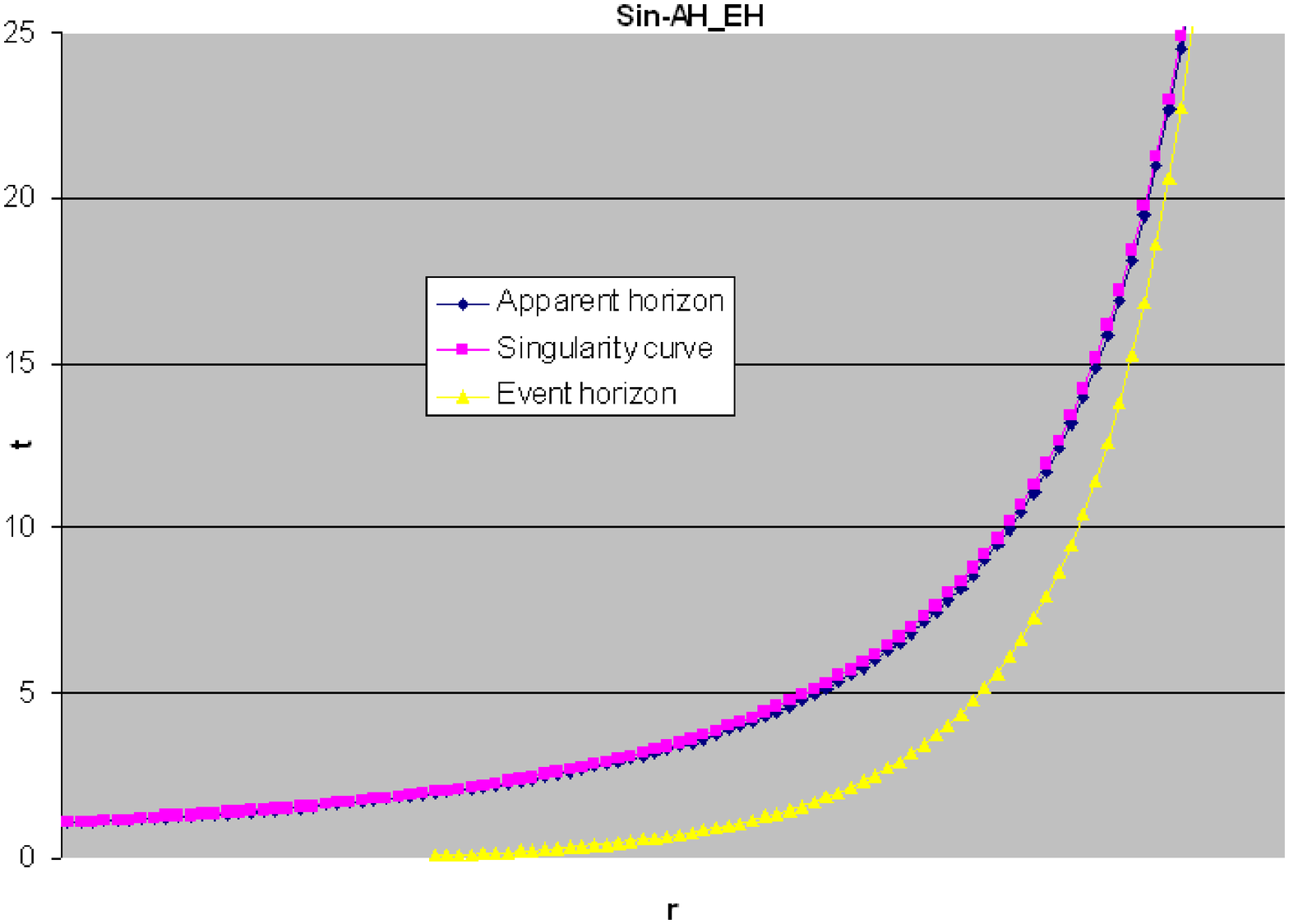}
\caption{ \label{den3} The case of asymptotically open model: the
density profile is similar to the previous open but asymptotically
flat case, except for the less mass concentrated in the central
region. The causal structure is also similar. The locations of the
event horizon, apparent horizon and the singularity are also
calculated numerically.The separation between the singularity and
the apparent horizon is not clear here due to the scale chosen.  }
\end{figure}

This solution represents a collapsing mass within an open FRW
universe. The collapsed region is again dynamical in the sense that
its mass is not constant, and the rate of change of the Misner-Sharp
energy is given by the same amount as the previous model. Therefore,
concepts of isolated horizon and slowly evolving horizon do not apply to this case.\\

\section{Characteristics of singularities of proposed models}

We have avoided in the models proposed the shell crossing
singularities except example III with a late time shell crossing
singularity.

The shell focusing singularities, however, are unavoidable and in
fact it is what we are looking for to study characteristics of
cosmological black holes. An important aspect of such a singularity
is its gravitational strength {\cite{strength},
which is an important differentiating feature of black hole.\\

\subsection{Strength of the shell focusing singularities}

 Heuristically, a singularity is termed gravitationally strong, or
simply strong, if it destroys by crushing or stretching any object
which falls into it. The prototype of such a singularity is the
Schwarzschild one: a radially infalling object is infinitely
stretched in the radial direction and crushed in the tangential
directions, with the net result of crushing to zero volume.
Otherwise a singularity is termed weak where no object falling into
it is destroyed. To check the strength of singularities of our
models we use the criteria defined by
Clarke {\cite{strength}. \\
Let $k^{\mu}$ be the tangent vector to the ingoing null geodesic,
and $\lambda$ the corresponding affine parameter being zero at the
center. $R_{\mu\nu}$ being the Ricci tensor, the singularity is said
to be strong if
 \begin{equation}
 \Psi=lim_{r\rightarrow 0}\lambda^{2}k^{\mu}k^{\nu}R_{\mu\nu} \neq 0.
\end{equation}

For a general LTB metric one obtains easily
$k^{\mu}k^{\nu}R_{\mu\nu}= 2(k^{t})^{2} \frac{M'}{R^{2}R'}$. For the
three interesting cases of cosmological black holes in flat and open
LTB models we have done the calculation along the lines of the
{\cite{joshi} using appropriate coordinates near singularity. For
our cases (\ref{met2-1}-\ref{met2-2}-\ref{met3}) we obtain after
some calculation $\Psi=0$ for $r\rightarrow 0$. Therefore, shell
focusing singularities occurring in the center of the models we are
proposing are week. This is in contrast to the Schwarzschild
singularity which is a strong one. We leave it to future studies if
this weakness is generic of any cosmological black holes.

\subsection{Nakedness of singularities}

We know already from Oppenheimer-Snyder collapse of a homogeneous
dust distribution how the shells become singular at the same time,
and thus none of them crosses. In the case of spherically symmetric
inhomogeneous matter configurations, however, the proper time of
collapse depends on the comoving radius $r$. Thus the piling up of
neighboring matter shells at finite proper radius can occur, thereby
producing two-dimensional caustics where the energy density and some
curvature components diverge. These singularities can be locally
naked, but they are gravitationally weak
{\cite{cencorship,nolan-sc-w}, i.e. curvature invariants and tidal
forces remain finite. It has also been shown that analytic
continuations of the metric, in a distributional sense, can always
be found in the neighborhood of the singularity {\cite{shell-c}.
\\
Models proposed in this paper are, however, free from shell crossing
singularities. The shell crossing singularity of example III at late
times does not influence the following argumentation. Conditions for
the absence of shell crossing singularities have been studied in
detail in {\cite{hell-shell}. In our case these conditions are
equivalent to $M'(r)>0$ and $R'>0$, which are satisfied by the
models discussed above. We may then conclude that

\begin{equation}
 \frac{\frac{dt}{dr}|_{AH}}{\frac{dt}{dr}|_{null}}=\left( 1 - \frac{2 M'}{R'} \right)<1.
\end{equation}
Therefore, the condition for the apparent horizon $R = 2M$ to be
spacelike is, i.e.
$-1<\frac{\frac{dt}{dr}|_{AH}}{\frac{dt}{dr}|_{null}}<1$, leads to
the condition $R'-M'>0$, which is not everywhere satisfied in our
model. As a result we notice that apparent horizons of the models
proposed here are not spacelike everywhere. Such a behavior has
already been discussed in{\cite{booth-mtt}.\\
 The case of shell focusing singularities is, however, a different one. Irrespective
of the behavior of the apparent and event horizons, it is then a
relevant question if the shell focusing singularity could be a naked
one. We notice that the slope of the outgoing null geodesics at the
singularity are greater than the slope of the singularity itself.
Therefore, the singularity is spacelike and no timelike or null
geodesic can come out of the singularity. We then conclude that the
singularities we are facing can not be naked.

\section{Discussion and conclusions}

Unlike models discussed so far in the literature, we have
constructed models of mass condensation within the FRW universe
leading to cosmological black holes without having the usual
pathologies we know from other models: the cosmic fluid is dust and
ideal producing a singularity at the center in the course of time.
The central singularity is spacelike and not naked. In the case of
flat or open universe models the singularity is weak and has
distinct apparent and event horizons. The apparent horizons are not
everywhere spacelike, to be compared with the Schwarzschild one
which is null everywhere. This has already been noticed in a general
context by {\cite{booth-mtt}. While the apparent horizon is defined
by the surfaces $R = 2M$, similar to the Schwarzschild horizon, the
even horizon is further away. Models we have proposed show that one
has to expect new effects while considering dynamical cosmic black
holes. The simple Schwarzschild static model may not reflect all the
phenomena one may expect in observational cosmology, and the black
hole thermodynamics. Even the simple concept of mass is not a
trivial one in such a dynamical environment. the answer to these
questions are beyond the scope of these paper and will be deal with
in future publications.

\appendix

  \section{}
 The curvature function $f(r)$ has a zero point where it changes sign for models III
and IV, corresponding to two different solutions. Therefore, we have
to take care of joining two solutions across the hypersurface
defined by $f(r) = 0$ to be continuous. This is done by looking at
the metric functions and their derivatives to be continuous.\\

Let us first look at the model IV. There we have to look at the
metric function $R$ and its derivatives, $R$, $R'$, $\dot{R}$,
$\ddot{R}$ and $\ddot{R}'$, at the point $r = 1$ where $f$ vanishes.
From the following relations derived from the Einstein equations
(\ref{ltbe00})
\begin{equation}
\ddot{R}=-\frac{M}{R^{2}},
 \end{equation}

\begin{equation}
\dot{R}'=\frac{M'}{R\dot{R}}-\frac{MR'}{\dot{R}R^{2}}+\frac{f'}{2\dot{R}},
 \end{equation}
and
 \begin{equation}
\ddot{R}'=-\frac{M'}{R^{2}}+\frac{2M R'}{R^{3}},
 \end{equation}
we infer that these second derivatives relevant for the Einstein
equations to be continuous on the hypersurface $f(r)= 0$ are
continuous if the $f$, $R$, $R'$, $\dot{R}$, $M'$, and $M$ are
continuous. Now, because of the continuity of $f$, $M'$, and $M$, we
just have to prove the continuity of $R$, $\dot{R}$, and $R'$. \\

Let us look first at $R$ and its derivative $R'$. In the case of
$r<1$ we have
\begin{eqnarray}
R=\frac{a(r)}{1-r}(1-cos\eta),\nonumber\\
\eta-sin\eta=\frac{(1-r)^{1.5}}{b(r)}t,
 \end{eqnarray}
where $a(r)=\sqrt{r}+r^{2}$, $b(r)=1+r^{1.5}$, and $a(1)=2$,
$a'(1)=2.5$, $b(1)=2$, $b'(1)=1.5$, and
\begin{equation}
\dot{R}=\frac{a \sqrt{1-r}}{b}\frac{sin\eta}{1-cos\eta}.
 \end{equation}
\begin{eqnarray}
R'=\frac{a'(1-r)+a}{(1-r)^2}(1-cos\eta)-\frac{a}{1-r}\nonumber\\
\frac{sin\eta}{1-cos\eta}\frac{1.5(1-r)^{0.5}b+b'(1-r)^{1.5}}{b^{2}}t.
 \end{eqnarray}
Defining $1-r=x$, we have
\begin{equation}
\eta-sin\eta=\frac{\eta^{3}}{6}-O(\eta^{5})=\frac{x^{3}}{2}t.
 \end{equation}
Therefore, to first order in $\eta$ we have
$\eta=\sqrt[3]{3t}\sqrt{x}$. Now taking the limit $x\rightarrow
0^{-}$ we obtain
\begin{eqnarray}
\lim_{x\rightarrow 0^{-}}R(x)= \lim_{x\rightarrow 0^{-}}
\frac{2}{x}(1-cos\eta)= \lim_{x\rightarrow
0^{-}}(\frac{\eta^{2}}{x}-O(\eta^{4})/x)\nonumber\\
= (3t)^{2/3}.~~~~~~~~~~~~~~~~~~~~~~~~
\end{eqnarray}
which is a well defined quantity.\\

In the case of $r > 1$ we have
\begin{eqnarray}
R=\frac{a(r)}{r-1}(cosh\eta-1),\nonumber\\
sinh\eta-\eta=\frac{(r-1)^{1.5}}{b(r)}t,
 \end{eqnarray}

\begin{equation}
\dot{R}=\frac{a \sqrt{r-1}}{b}\frac{sinh\eta}{cosh\eta-1},
 \end{equation}
and

\begin{eqnarray}
R'=\frac{a'(r-1)-a}{(r-1)^2}(cosh\eta-1)+\frac{a}{r-1}\nonumber\\
\frac{sinh\eta}{cosh\eta-1}\frac{1.5(r-1)^{0.5}b-b'(r-1)^{1.5}}{b^{2}}t.
 \end{eqnarray}

 Now, defining $r - 1=x$, and noting that
\begin{equation}
sinh\eta-\eta=\frac{\eta^{3}}{6}+O(\eta^{5})=\frac{x^{3}}{2}t,
 \end{equation}
we obtain to first order of $\eta$ the relation
$\eta=\sqrt[3]{3t}\sqrt{x}$. Therefore,
\begin{eqnarray}
\lim_{x\rightarrow 0^{+}}R(x)= \lim_{x\rightarrow 0^{+}}
\frac{2}{x}(cosh\eta-1)= \lim_{x\rightarrow
0^{+}}\frac{\eta^{2}}{x}+O(\eta^{4})/x\nonumber\\
=(3t)^{2/3}~~~~~~~~~~~~~~~~~~~~~~~~~~~~.
\end{eqnarray}
Therefore the continuity of $R$ across $r = 1$ is established.\\
Similar calculation for the first derivatives shows the continuity
of $R'$ and $\dot{R}$ having well defined values on both sides of
the $r=r_{0}$ hypersurface:

\begin{equation}
R'(1)=2.5(3t)^{2/3}-\frac{3}{4(3t)^{1/3}} t,
 \end{equation}
and

\begin{equation}
\dot{R}(1)=\frac{2}{(3t)^{1/3}} .
 \end{equation}

The case of model III is similar except for the hypersurface defined
by $g(r)=\frac{f(r)}{r}=e^{-r}-\frac{1}{r^{2}+20000} = 0$ with the
root of $e^{-r_{0}}-\frac{1}{r_{0}^{2}+20000}=0$ being at a point $r
= r_0$ different from  $r=1$. It is easy to see that $g(r)$ is an
analytic function at $r=r_{0}$, and can be approximated by
$g(r)\approx
g'(r_{0})(r-r_{0})+\frac{g''(r_{0})}{2}(r-r_{0})^{2}+...$. Similar
calculations verify the continuity of the metric function $R$ and
its relevant derivatives across the hypersurface $r=r_0$.


\begin{thebibliography}{}
%
\bibitem{sultana}
J. Sultana and C.C. Dyer, Gen. Rel. Grav. 37, 1349 (2005).

\bibitem{Einstein Straus}
A. Einstein and E.G. Straus, Rev. Mod. Phys. 17, 120 (1945); 18, 148
(1946).

\bibitem{Baker}
G.A. Baker Jr., astro-ph/0003152.
\bibitem{nolan m}
B.C. Nolan, J. Math. Phys. 34, 1 (1993).
\bibitem{McVittie}
G.C. McVittie, Mon. Not. R. Astr. Soc. 93, 325 (1933).
\bibitem{haw-bh}
Hawking S W, Ellis G F R,  \emph{The Large Scale Structure of
Space-Time} (Cambridge University Press, 1973).
\bibitem{doeleman}
Doeleman, S. S. et al. Nature 455, 78–80 (2008).
\bibitem{Hayward94}
S. A. Hayward, Phys. Rev. D 49, 6467 (1994).
\bibitem{ashtekar99}
A. Ashtekar, C. Beetle, O. Dreyer, S. Fairhurst, B. Krishnan, J.
Lewandowski and J. Wisniewski, Phys. Rev. Lett. 85, 3564-3567
(2000).
\bibitem{ashtekar02}
A. Ashtekar and B. Krishnan, Phys. Rev. Lett. 89, 261101 (2002).
\bibitem{booth04}
Booth I. and Fairhurst S., Phys. Rev. Lett. 92, 011102 (2004).
\bibitem{Eardley98}
D. Eardley, Phys. Rev. D 57, 2299 (1998).
\bibitem{Krishnan05}
Erik Schnetter and Badri Krishnan, Phys. Rev. D 73, 021502, (2006).
\bibitem{Ben-Dov}
Ishai Ben-Dov, Phys. Rev. D 75, 064007, (2007).
\bibitem{harada}
Tomohiro Harada, C. Goymer and B.J. Carr,  Phys. Rev. D66, 104023
(2002)

\bibitem{mis-sharp}
Misner C W and Sharp D H 1964 Phys. Rev. 136 B571.
\bibitem{LTB}
R. C. Tolman, Proc. Natl. Acad. Sci. U.S.A. 20, 410 (1934); H.
Bondi, Mon. Not. R. Astron. Soc. 107, 343 (1947); G. Lemaiˆ- tre,
Ann. Soc. Sci. Bruxelles I A53, 51 (1933).


\bibitem{cencorship}
P. S. Joshi and I. H. Dwivedi, Phys. Rev. D 47, 5357 (1993); R. P.
A. C. Newman, Class. Quantum Grav. 3, 527 (1986); D. Christodoulou,
Commun. Math. Phys. 93, 171 (1984); D. M. Eardly and L. Smarr, Phys.
Rev. D 19, 2239 (1979).
\bibitem{joshi} P. S. Joshi and T. P. Singh, Phys. Rev. D 51,
6778 (1995).
\bibitem{joshi-cell}
A. Chamorro, S.S. Deshingkar, I.H. Dwivedi, and P.S. Joshi, Phys.
Rev. D 63, 084018 (2001).
\bibitem{initial condition}
 I. H.
Dwivedi and P. S. Joshi, Class. Quantum Grav. 14, 1223 (1997); S. S.
Deshingkar, S. Jhingan, and P. S. Joshi, Gen. Relativ. Gravit. 30,
1477 (1998).
\bibitem{mansouri}
Khakshournia and R. Mansouri, Phys. Rev. D 65, 027302, (2002); V.A.
Berezin, V.A. Kuzmin, and I.I. Tkachev, Phys. Rev. D 36, 2919
(1987).
\bibitem{rindler}
W. Rindler, M. Ishak, Phys. Rev. D 76 043006 (2007).
\bibitem{wald98}
R. M. Wald, (gr-qc/ 9710068).
\bibitem{kra-hel-sf}
A. Krasin´ski and C. Hellaby, Phys. Rev. D 65, 023501 (2002); C.
Hellaby and A. Krasin´ski , Phys. Rev. D 73, 023518 (2004) .

\bibitem{kra-hel-BH}
A. Krasin´ski and C. Hellaby, Phys. Rev. D 69, 023502 (2004);


\bibitem{faraoni2}
V. Faraoni and A. Jacques,Phys Rev D78, 024008 (2008).

\bibitem{nolan-sc-w}
B. C. Nolan, Phys. Rev. D 60, 02014 (1999).
\bibitem{shell-c}
P. Yodzis, H. J. Seifert, and H. M. zum Hagen, Comm. Math. Phys. 34,
135 (1973); A. Papapetrou and Hamoui, Ann. Inst. Henr´ý Poincar´e
VI, 343 (1967).

\bibitem{origin con}
N. Mustapha and C. Hellaby, Gen. Rel. Grav. 33, 455-77 (2001).
\bibitem{strength}
C. J. S. Clarke, Class. Quant. Grav. 10, 1375 (1993); F. J. Tipler,
Phys. Lett. A 64A, 8 (1977); C. J. S. Clarke and A. Krolak, J. Geom.
Phys. 2, 127 (1985).

\bibitem{Einstein Straus1}
W.B. Bonnor, Class. Quant. Grav. 16, 1313 (1999); J.M.M. Senovilla
and R. Vera, Phys. Rev. Lett. 78, 2284 (1997); M. Mars, Phys. Rev. D
57, 3389 (1998).


\bibitem{na73}
Demia´nski M. and Lasota JP. Nature Phys. Sci. 241, 53 (1973).
\bibitem{tipler77}
Frank J. Tipler. Nature 270, 500 (1977).
\bibitem{nolan}
B. C. Nolan, Class. Quantum Grav. 16, 1227 (1999); B. C. Nolan,
Phys. Rev. D 58, 064006 (1998); Class. Quantum Grav. 16, 3183
(1999).

\bibitem{booth-mtt}
I. Booth, L. Brits, J. A. Gonzalez, and C. Van Den Broeck. Class.
Quant. Grav. 23 413-440 (2006).

\bibitem{hell-shell}
C. Hellaby and K. Lake, Astrophys. J. 290, 381 (1985); 300, 461(E)
(1985).



\end{thebibliography}
\end{document}